\documentclass[a4paper,10pt]{article}\usepackage{graphics}
\usepackage{epsfig}
\usepackage{epstopdf}
\usepackage{amsmath}
\title{\textbf{Cosmological horizon entropy and generalised second law for flat Friedmann universe}}
\author{Titus K Mathew$^1$, 
Aiswarya R$^2$  Vidya K Soman$^3$ \\ 
Department of Physics, \\ Cochin University of Science and Technology, \\ Kochi-22, India. \\
E-mail:$^1$titus@cusat.ac.in, $^2$aiswaryar.r95@gmail.com, \\ $^3$vidyaks10@gmail.com}
\date{}
\begin{document}

\maketitle

\begin{abstract}
We discuss the generalized second law (GSL) and the constraints imposed by it for two types of Friedmann universes. The 
first one is the Friedmann universe with radiation and a positive cosmological constant, and the second one consists of 
non-relativistic matter and a positive cosmological constant. The time evolution of the event horizon entropy and the entropy of the 
contents within the horizon are analyses in an analytical way by obtaining the Hubble parameter. It is shown that the GSL constraint 
the temperature of both the radiation and matter of the Friedmann universe. It 
is also shown that, even though the net entropy of the radiation (or matter) is decreasing at sufficiently large times 
as the universe expand, it exhibit an increase during the early times when universe is decelerating. That is the entropy 
of the radiation within the comoving volume is decreasing only when the universe has got an event horizon.
\end{abstract}


\vspace{0.2in}


\section{Introduction}
\label{intro}
Bekenstein and Hawking have showed that the entropy of black holes is proportional to the
area of their event horizon \cite{Beken1,Hawking1,Bardeen1}. In units of $C=1, G=1, k=1$ and $\hbar=1$, the black hole
entropy is given as
\begin{equation} \label{eqn:hentro}
 S = {A_h \over 4}
\end{equation}
where $A_h$ is the area of event horizon of the black hole. Hawking have shown that the black hole can evaporate
by emitting radiation, consequently  it's event horizon area decreases. He had also shown that the event horizon 
of the black hole
posses temperature, which is inversely proportional to it's mass or proportional to it's surface gravity. 
During the process of evaporation the entropy
of the black hole will decrease. But due to the emitted radiation, the entropy of the surrounding universe will 
increase. Hence the second law of thermodynamics was modified in such a way that, 
the entropy of the black hole plus the entropy of the exterior environment of the black hole will never decrease, this is
called as the generalized second law(GSL), which can be represented as,
\begin{equation}
 {d \over dt}(S_{evn} + S_b) \geq 0
\end{equation}
where $S_{env}$ is the entropy of environment exterior to the black hole and $S_b$ is the entropy of the black hole.

 The thermodynamic properties of the event horizon, was shown to exist 
in a more basic level\cite{Paddy1,Paddy2}, by recasting the Einstein's field equation for a spherically symmetric space time as
in the form of the first law of thermodynamics. In references \cite{Cai1,Cai2} one can find investigations on the applicability 
of the first law of thermodynamics to cosmological event horizon. Jacobson \cite{Jacob1} showed that, Einstein's field equations are equivalent to the 
thermodynamical equation of state of the space time.

In cosmology the counter part of black hole horizon is the cosmological event horizon. Gibbons and Hawking \cite{Gibbons1} proposed that
analogous to black hole horizon, the cosmological event horizon also do possess entropy, proportional to their area.
They have proved it particularly for de Sitter universe for which an event horizon is existing. For cosmological horizon, GSL 
implies that, the entropy of the horizon together with the matter enclosed by the event horizon of the universe will never decrease.
That is  the rate of change
of entropy of the cosmological event horizon together with that of material contents within the horizon of the universe, 
must be greater than or 
equal to zero,
\begin{equation} \label{eqn:gsl}
 {d \over dt}(S_{CEH} + S_{m}) \geq 0
\end{equation}
where $S_{CEH}$ is the entropy of the cosmological event horizon and $S_m$ represents the entropy of the matter or radiation (or both 
together) of the universe. The validity of GSL for cosmological horizon was confirmed and extended to universe consisting of radiation 
by numerical analyses by  
Davies \cite{Davies1} and others \cite{Pollock1,Brustein1,Diego1,Wang1,Sergio1}. In reference \cite{Karami1,Rahul1}, the authors analyzed the 
GSL with some variable models of f(T) gravity. In reference \cite{Nairwita1} GSL was analyzed with reference to brane scenario. 
Ujjal Debnath et. al. \cite{Ujjal1} have analyzed the GSL for FRW cosmology with power-law entropy correction.

There are investigations connecting the entropy and hidden information. In the case of black hole horizon, the observer is outside
the horizon, and the entropy of the black hole is considered as measure of the information hidden within the black hole.
 While regarding cosmological horizon, 
the observer is inside the horizon. This will cause problems in explaining the 
entropy of the cosmological horizon as the measure of hidden information as in the case of black hole. In the case of 
black hole the hidden region is finite, while in the case of cosmological horizon, there may be infinite region 
beyond the event 
horizon of the universe, which  causing problems in explaining the cosmological horizon entropy as the 
hidden information.

Another important fact is regarding the existence of dominant energy condition for the non decreasing horizon area. 
In the case of black hole, Hawking proved an area theorem, that the area of the black hole will never decrease if it is not 
radiating \cite{Hawking2}. Davies \cite{Davies3} proved an analogous theorem for cosmological event horizon that the area of the cosmological
event horizon 
will never decrease, provided it satisfies the dominant energy condition,
\begin{equation} \label{eqn:condition1}
 \rho + p \geq o
\end{equation}
where $\rho$ is the 
density of the cosmic fluid and $p$ is its pressure. 

Regarding the applicability of the generalized second law to the Friedmann universe, analysis were done by considering
 the Friedmann universe as a small deviation form the de Sitter phase\cite{Davies2,Davies3}. 
In these works the authors calculated the horizon entropy through a numerical computation of the comoving distance to the 
event horizon. 
In the present work we 
obtained an analytical equation for the Hubble parameter and proceeded to the calculation of the entropy of the cosmological 
event horizon in an analytical way. We also checked the validity of dominant energy condition by using the 
derived Hubble parameter. Our analysis is 
for a flat
universe which consists of (i) radiation and positive cosmological constant and (ii) non-relativistic matter and positive 
cosmological constant. We have considered the flat universe because of the fact that, the inflationary cosmological models 
predicts flat universe and more over the flatness of the space is confirmed by observations, for example, the current value of the 
curvature parameter is $\Omega_{k0} \sim 10^{-3}$ \cite{Komat}.

The paper is arranged as follows. In section two, we consider the flat Friedmann universe with radiation and a positive cosmological 
constant. We are presenting the calculation of the entropy of radiation, event horizon and the total entropy of universe and the 
respective time evolutions. We have also checked the validity of the generalized second law in this section. In section three 
we present the analogous 
calculations for the flat Friedmann universe with non-relativistic matter and a positive cosmological constant. In section four we
present the particular behaviour of the radiation entropy in the Friedmann universe with reference to the development 
of the event horizon. In section five we present the discussion followed by conclusions.

\section{Friedmann universe with radiation and cosmological constant.}
\label{sec:1}
For a flat Friedmann universe with FRW metric, the dynamics are governed by the Friedmann equations(by choosing $8\pi G =1$),
\begin{equation} \label{eqn:friedmann1}
 H^2 = {\rho_{\gamma} \over 3} + {\Lambda \over 3}
\end{equation}
and
\begin{equation} \label{eqn:friedmann2}
 \dot{\rho_{\gamma}} + 3H \left(\rho_{\gamma} + P_{\gamma} \right) = 0
\end{equation}
where $\rho_{\gamma}$ is the radiation density, $p_{\gamma}$ is the radiation pressure, $\Lambda$ is the constant cosmological 
constant, $H$ is the Hubble parameter and the dot over the density represents derivative with respect to time. For 
radiation, the pressure is, $p_{\gamma} = \rho_{\gamma} /3$. From equations 
(\ref{eqn:friedmann1}) and  (\ref{eqn:friedmann2}) the scale factor of this universe can be obtained as, 
\begin{equation} \label{eqn:afact1}
 a(t) = \left({\Omega_{\gamma 0} \over \Omega_{\Lambda}} \right)^{1/4} \sinh^{1/2}(2 \sqrt{\Omega_{\Lambda}} H_0 (t-t_0))
\end{equation}
where $\Omega_{\gamma 0} = \rho_{\gamma 0} /3H_0^2$, $\Omega_{\Lambda} = \Lambda/3H_0^2$ and $H_0$ is the present value of 
the Hubble parameter.
This equation shows that as $t \rightarrow t_0$ the scale factor $a(t) \rightarrow t^{1/2}$ the radiation dominated phase of the 
Friedmann universe and as $t \rightarrow \infty$ the scale factor $\displaystyle a(t) \rightarrow e^{\sqrt{\Lambda \over 3} t} $ , 
\begin{figure}
\includegraphics[scale=0.75]{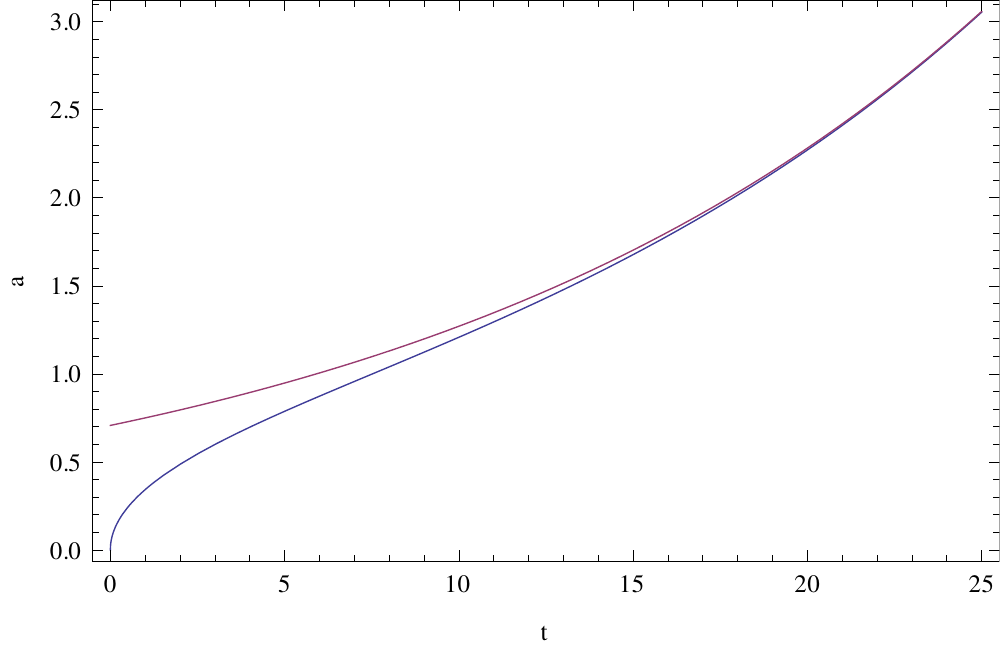}
\caption{ \emph{Evolution of scale factor $a(t)$ with time $t$ in Gyrs, the blue line (lower line) corresponds to the Friedmann 
universe and red line (upper line) for de Sitter universe.}}
\label{fig:af1}
\end{figure} 
the de Sitter phase. The behaviour of the scale factor with time is shown in figure \ref{fig:af1}, in comparison with the scale 
factor of the de Sitter phase. From the plot it is evident that the scale factor of the Friedmann 
universe tends to the de Sitter phase at large times. So at smaller times the universe is in the radiation dominated phase and it is 
decelerating, consequently it doesn't have event horizon. At larger times the universe enters the accelerated expansion phase, 
where it posses an event horizon.

\begin{figure}
 \label{fig:codist1}
\includegraphics[scale=0.75]{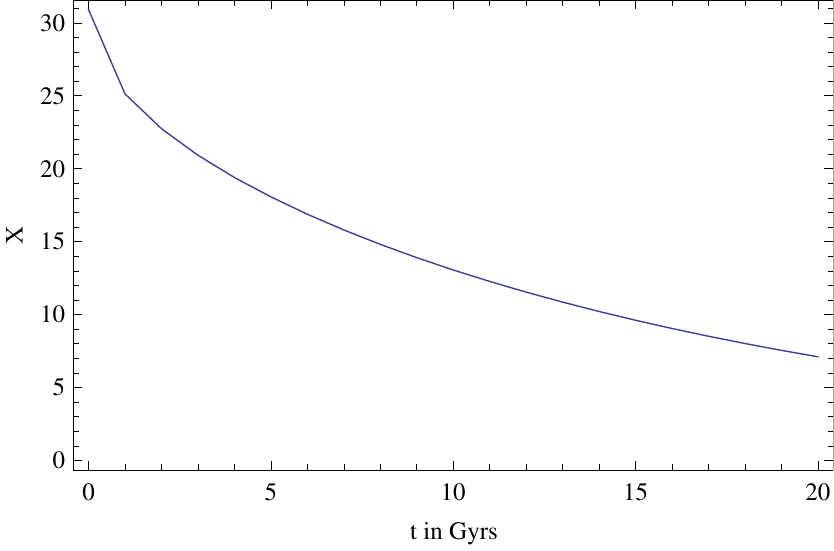}
\caption{Variation of the comoving distance $\chi$ to the event horizon with time in Gyrs for Friedmann universe with radiation 
and cosmological constant}
\end{figure}

The co-moving distance to the event horizon, can be obtained by using the relation,
\begin{equation} \label{eqn:chi}
 \chi = \int_t^{\infty} {dt \over a(t)} < \infty
\end{equation}
Thus the proper distance to the event horizon is $ r_h = a(t) \chi.$ For the existence of the event horizon, the integral has to converge. 
With the scale factor in equation (\ref{eqn:afact1}), the integral in the equation for comoving distance to the event horizon cannot be 
solved analytically.
So as a first step we made a numerical 
computation of the comoving distance to the event horizon, as it is necessary to understand the time evolution of the comoving 
distance and the result is shown in figure \ref{fig:codist1}. 
The plot shows that the comoving distance to the event horizon is decreasing with time. Since
the comoving horizon distance is decreasing, the comoving volume of the universe within the horizon also decreases. 
The radiation density behaves as $\rho_{\gamma} \sim a^{-4}$, therefore the radiation content within the horizon is 
decreasing with time. Which nevertheless implies that the radiation is crossing the horizon, hence the radiation 
entropy within the horizon is decreasing. This method and conclusion is in line with the result of T M Davies et. al. 
\cite{Davies2}. One can also find investigations of the same spirit regarding the heat flow through the cosmic horizon in 
references \cite{Bousso1,Frolov1}. In fact this result is true for any model of the universe having an event horizon.

The horizon entropy can be obtained as per the Bekenstein equation (\ref{eqn:hentro}). For that the area of the event horizon 
can be taken 
as $A_h= 4\pi \chi^2.$   In the work of Davies et, al. the entropy was calculated in a numerical way, but we are obtaining the entropy 
of the horizon using the Hubble parameter obtained form the scale factor. We are substituting $\chi$ in terms of the 
Hubble parameter. The scale factor in equation (\ref{eqn:afact1}) 
shows that, at large time the scale factor is
approaching to that of de Sitter phase. For de Sitter phase, it can be shown that,
\begin{equation}
 \chi = {1 \over H}
\end{equation}
Since the Friedmann universe considered here is approaching the de Sitter phase at large times, it will not be unfair in taking, the 
comoving distance $ \chi \sim 1/H $ for the Friedmann universe in consideration. For the scale factor in equation (\ref{eqn:afact1}), 
the Hubble parameter is,
\begin{equation} \label{eqn:H1}
 H = H_0\sqrt{\Omega_{\Lambda}} \coth(2\sqrt{\Omega_{\Lambda}}H_0 t)
\end{equation}

\begin{figure}[ht]
\includegraphics[scale=0.75]{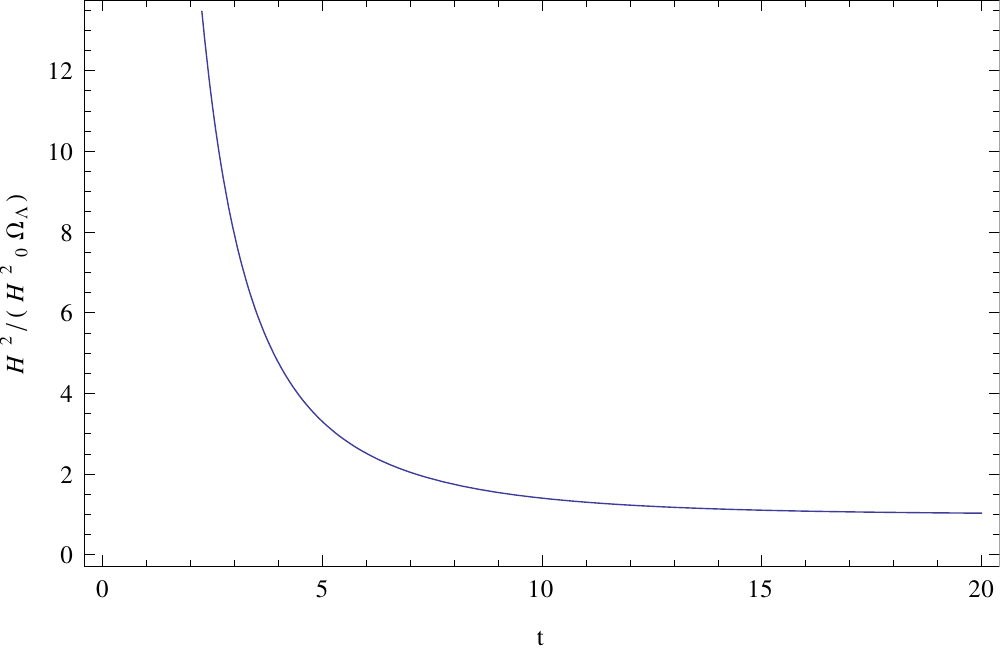}
\caption{the plot corresponds to the non-decreasing condition of the area of event horizon, equation (\ref{eqn:condition1})}
\label{fig:cond1}
\end{figure}

Before going for a calculation of the entropy of the event horizon, we will check here the validity of the area theorem proposed 
by Davies, with the obtained Hubble parameter.
From equation (\ref{eqn:friedmann1}) and (\ref{eqn:H1}), the condition for non-decreasing horizon area, 
equation (\ref{eqn:condition1}), leads to
\begin{equation}
 {H^2 \over H_0^2 \Omega_{\Lambda}} \geq 1
\end{equation}
Using equation (\ref{eqn:H1}) we have plotted $\displaystyle {H^2 \over H_0^2 \Omega_{\Lambda}}$ versus time in figure \ref{fig:cond1}.
 We have used the 
parameter values, $H_0=73 \, km \, s^{-1} \, Mpc^{-1}$ \cite{Amanullah1} and a standard value $\Omega_{\Lambda} = 0.7$ through out for our
calculations. The plot shows that the area of the
event horizon of the Friedmann universe with radiation and a positive cosmological constant will never decrease, hence the 
entropy of horizon will never decrease.
On the other hand, the entropy of the radiation is decreasing with time as we have argued earlier.

In oder to satisfy the GSL, the decrease in the entropy of the radiation is to be balanced by the increase in the 
horizon entropy. The horizon area is always increasing, implies that there exist some kind of trading of the entropy between the 
horizon and the radiation content of the universe. The entropy of the event horizon is $\displaystyle S_{CEH} = \pi \chi^2.$
As we have argued earlier, taking $\chi = 1/H$, the horizon entropy become,
\begin{equation} \label{eqn:CEH2}
S_{CEH} = {\pi \over H^2} = {\pi \over H_0^2 \Omega_{\Lambda} \coth^2(2 \sqrt{\Omega_{\Lambda}} H_0 t)}
\end{equation}
The entropy of the radiation can be obtained using the relation,
\begin{equation} \label{eqn:radentro}
 S_{\gamma} = \left( {{\rho_{\gamma} + p_{\gamma}} \over T_{\gamma}} \right) V_{CEH}
\end{equation}
where $V_{CEH}$ is the volume of the event horizon and $T_{\gamma}$ is the temperature of the radiation. Taking $V_{CEH} = 4 \pi \chi^3 /3$, substituting
temperature form radiation
energy density, $\displaystyle \rho_{\gamma} = \sigma T_{\gamma}^4$, 
\begin{equation} \label{eqn:sgamma1}
 S_{\gamma} = {16 \pi \sigma^{1/4} \over 9} {\rho_{\gamma}^{3/4} \over H^3}
\end{equation}
which after substituting 
$H$ parameter form equation (\ref{eqn:H1}) and $\rho_{\gamma}$ in terms of $H$, using the Friedmann equations,
become
\begin{equation} \label{eqn:sgamma2}
 S_{\gamma} = {16 \pi \sigma^{1/4} \over 9} {\left(3 H_0^2 \Omega_{\Lambda} \coth^2(2\sqrt{\Omega_{\Lambda}}H_0 t) -
 3 H_0^2 \Omega_{\Lambda} \right)^{3/4} \over \left( H_0 \sqrt{\Omega_{\Lambda}} \coth(2\sqrt{\Omega_{\Lambda}}H_0 t)\right)^3 }
\end{equation}
where  $\sigma = \pi^2 /15,$ the radiation constant. 
\begin{figure} 
\includegraphics[scale=0.75]{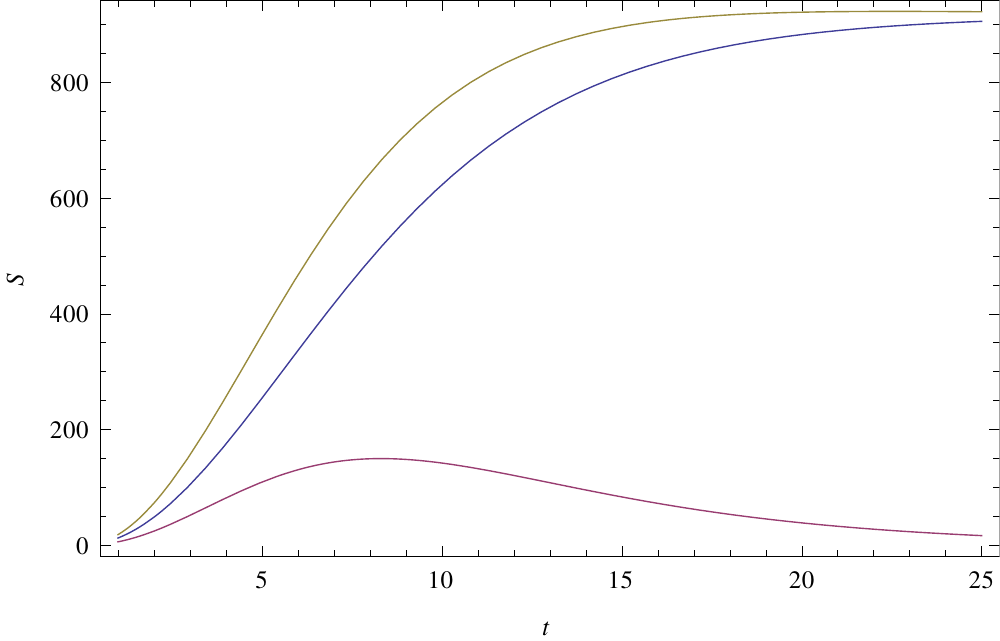}
\caption{The variation of entropies with time in Gigayears. The lower line representing the time evolution of radiation entropy, 
the top most line is that corresponds to
event horizon and the middle is for the sum of horizon and radiation entropies.}
\label{fig:entrorad}
\end{figure}

We have plotted the time variation of $S_{CEH}$, $S_{\gamma}$ and $S_{CEH}+S_{\gamma}$ in figure 4. The figure shows that, at sufficiently 
large times the radiation entropy is decreasing, while the horizon entropy is increasing. The increase 
in the horizon entropy is more than required to compensate for the decrease in the radiation entropy because of that,
total entropy 
comprising the entropy of the radiation and horizon will increase. This is confirming the validity of the 
generalized second law for the cosmological horizon, that the entropy of the horizon plus the entropy of the fluid within the 
horizon will never decrease. This result is in confirmation with the earlier works of Davies and others, but they have
arrived at the conclusion through straight numerical work, on the other hand our work is more of an analytical way.

The conditions for satisfying the generalized second law can be obtained by analysing the validity of the exact statement of the 
law as given equation (\ref{eqn:gsl}).
The time rate of the horizon entropy is,
\begin{equation}
{d S_{CEH} \over dt} = - 2 \pi \left({\dot{H} \over H^3} \right) 
\end{equation}
where the dot over $H$ represents the derivative with time given as $ \displaystyle \dot{H} = -(2 \rho_{\gamma} / 3)$, leads to
\begin{equation}
 {d S_{CEH} \over dt} = {4 \pi \rho_{\gamma} \over 3 H^3}.
\end{equation}
 The time rate of radiation entropy can be given from (\ref{eqn:sgamma1})
as,
\begin{equation} \label{eqn:radentrot}
 {d S_{\gamma} \over dt} = - {16 \pi \sigma^{1/4} \rho_{\gamma}^{3/4} \over 3}   \left( {1 \over H^2} + {\dot{H} \over H^4} \right)
\end{equation}
The above two equations reveal that the time rate of horizon entropy is positive hence the horizon entropy is at the increase,
while the time rate of radiation entropy is negative hence the radiation entropy is at the decrease. The generalized second law,
can then be represented as,
\begin{equation} \label{eqn:gslcon3}
 - 2 \pi \left({\dot{H} \over H^3} \right) - {16 \pi \sigma^{1/4} \rho_{\gamma}^{3/4} \over 3}   \left( {1 \over H^2} + 
{\dot{H} \over H^4} \right) \geq 0
\end{equation} \label{eqn:gslcond3}
 Replacing $H$ and $\dot{H}$ using the equation (\ref{eqn:H1}), the above condition become,
\begin{equation} \label{eqn:gsl3}
\begin{split}
 {H_0 \sqrt{\Omega_{\Lambda}} cosech^2(2\sqrt{\Omega_{\Lambda}} H_0 t) \over coth(2\sqrt{\Omega_{\Lambda}} H_0 t)}
- {4 \sigma^{1/4} \rho_{\gamma}^{3/4} \over 3}  \\ \left(1 - 2 sech^2(2\sqrt{\Omega_{\Lambda}} H_0 t) \right) \geq 0
\end{split}
\end{equation}

\begin{figure}
\includegraphics[scale=0.75]{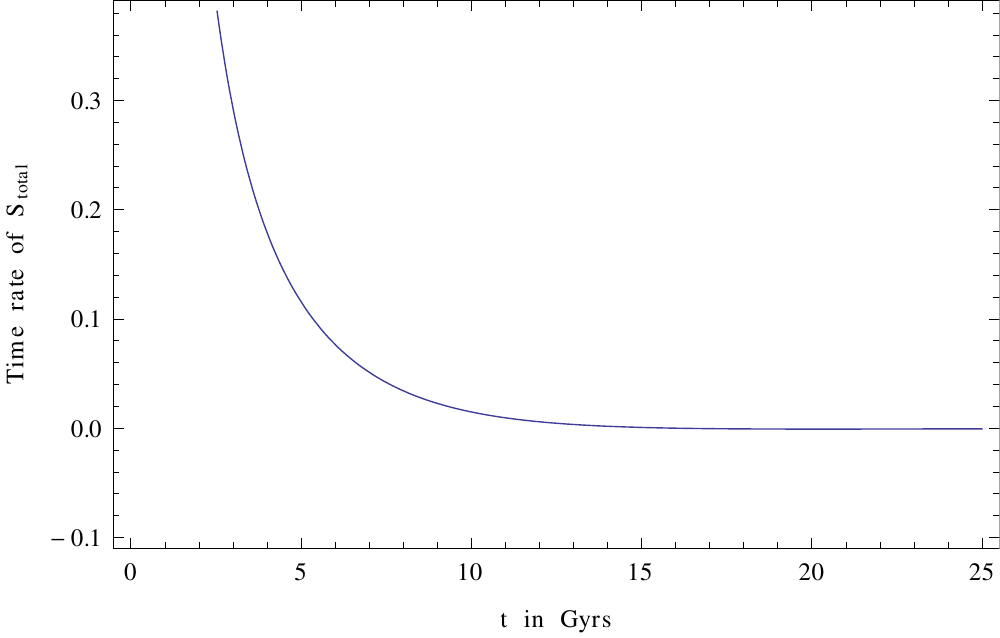}
\caption{Plot of the left hand side of the GSL condition in equation \ref{eqn:gsl3}} 
\label{fig:gslcon1}
\end{figure}

Expressing $\rho_{\gamma}$, in terms of $H$, using the Friedmann equation, we have evaluated the time evolution of the 
left hand side of
the above inequality condition and the result is shown in figure \ref{fig:gslcon1}. 
The figure shows that the condition for the GSL
is always satisfied.

From the GSL condition in equation(\ref{eqn:gslcon3}), we can obtain a condition regarding the temperature of the radiation 
within the horizon. With $\dot{H}=-2\rho_{\gamma}/3$, the condition (\ref{eqn:gslcon3}), become,
\begin{equation}
 {\rho_{\gamma}^{1/4} \over H} + 8\sigma^{1/4}{\rho_{\gamma} \over 3 H} \geq 4 \sigma^{1/4}
\end{equation}
Taking $\rho_{\gamma}^{1/4} = \sigma^{1/4} T_{\gamma}$, then an inequality condition constraints the present value of the temperature of 
the radiation can be obtained as,
\begin{equation}
 T_{\gamma 0} \geq \left(4 H_0 - 8 H_0 \Omega_{\gamma 0} \right)
\end{equation}
the above condition leads to a numerical value, $T_{\gamma 0} \geq 10^{-29}$K.
Compared to the present temperature of the 
radiation $T_{\gamma 0} \sim 2.73 K$, this is very much in favour of the validity of second law in the Friedmann universe.
This result is agreeing with the result obtained by Davies et. al, that $T_{\gamma 0} \geq H$, where $H$ is now 
taken as the temperature of the horizon. By using the fundamental constants, the temeprature of the horizon, is
\begin{equation}
 T_{CEH} = {\hbar \over k} {H \over 2\pi}
\end{equation}
where $k$ is the Boltzmann constant, which implies a present value, $T_{CEH 0} \sim 10^{-30}$K. So the temperature of the horizon
is less than that of the event horizon, which indicates that, the radiation can aproach the horizon.
   
\section{Freidmann universe with matter and cosmological constant.}

In this section we are analysing the Friedmann universe with matter and a positive cosmological constant, regarding the 
horizon entropy and the generalized second law. The Friedmann equation, in this case is,
\begin{equation}
 H^2 = {\rho_m \over 3} + {\Lambda \over 3}
\end{equation}
The scale factor can then be obtained as,
\begin{equation} \label{eqn:a2}
 a(t) = \left({\Omega_{m0} \over \Omega_{\Lambda} }\right)^{1/3} \sinh^{2/3}({3\sqrt{\Omega_{\Lambda}} H_0 t \over 2})
\end{equation}
As in the previous case, here also the solution will tends to the de Sitter phase, $\displaystyle a(t) \rightarrow e^{\sqrt{\Lambda/3}}$ 
as $t \rightarrow \infty.$ Which means that the model posses an event horizon. 
The Hubble parameter corresponds to the scale factor is,
\begin{equation} \label{eqn:H2}
 H = H_0 \sqrt{\Omega_{\Lambda}} \coth(\frac{3}{2}\sqrt{\Omega_{\Lambda}} H_0 t)
\end{equation} 
It can be seen that the dominant energy condition is being satisfied, as in the case of Friedmann universe with radiation, such that
$\displaystyle H^2/(H_0^2 \Omega_{\Lambda}) \geq 1$
at all time.

\begin{figure}
 \includegraphics[scale=0.7]{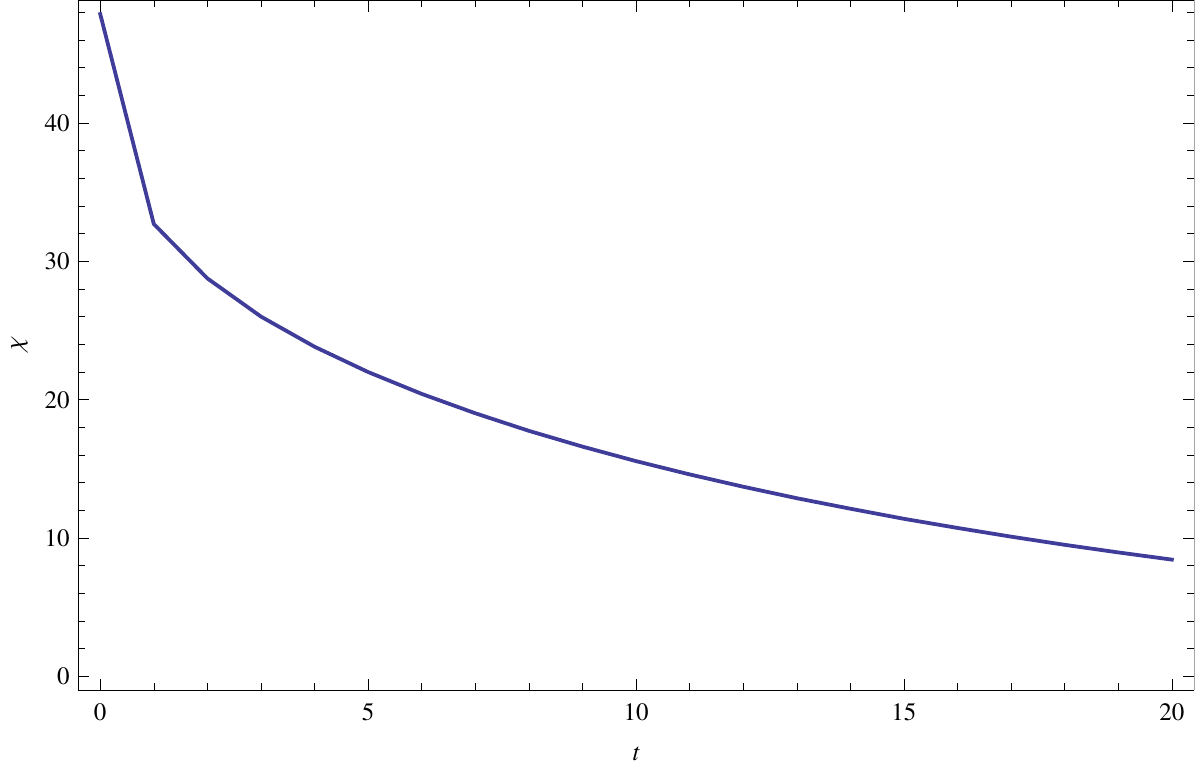}
\caption{Comoving distance to the horizon of Friedman universe with matter and $\Lambda$}
\label{fig:dist1}
\end{figure}

The comoving distance to the horizon is evaluated using the scale factor in the equation (\ref{eqn:a2}), and is deceasing 
with time as shown is shown in figure
\ref{fig:dist1}. 
So the matter entropy within the horizon
will decrease and hence the matter will crosses the event horizon. 

\begin{figure}
 \includegraphics[scale=0.7]{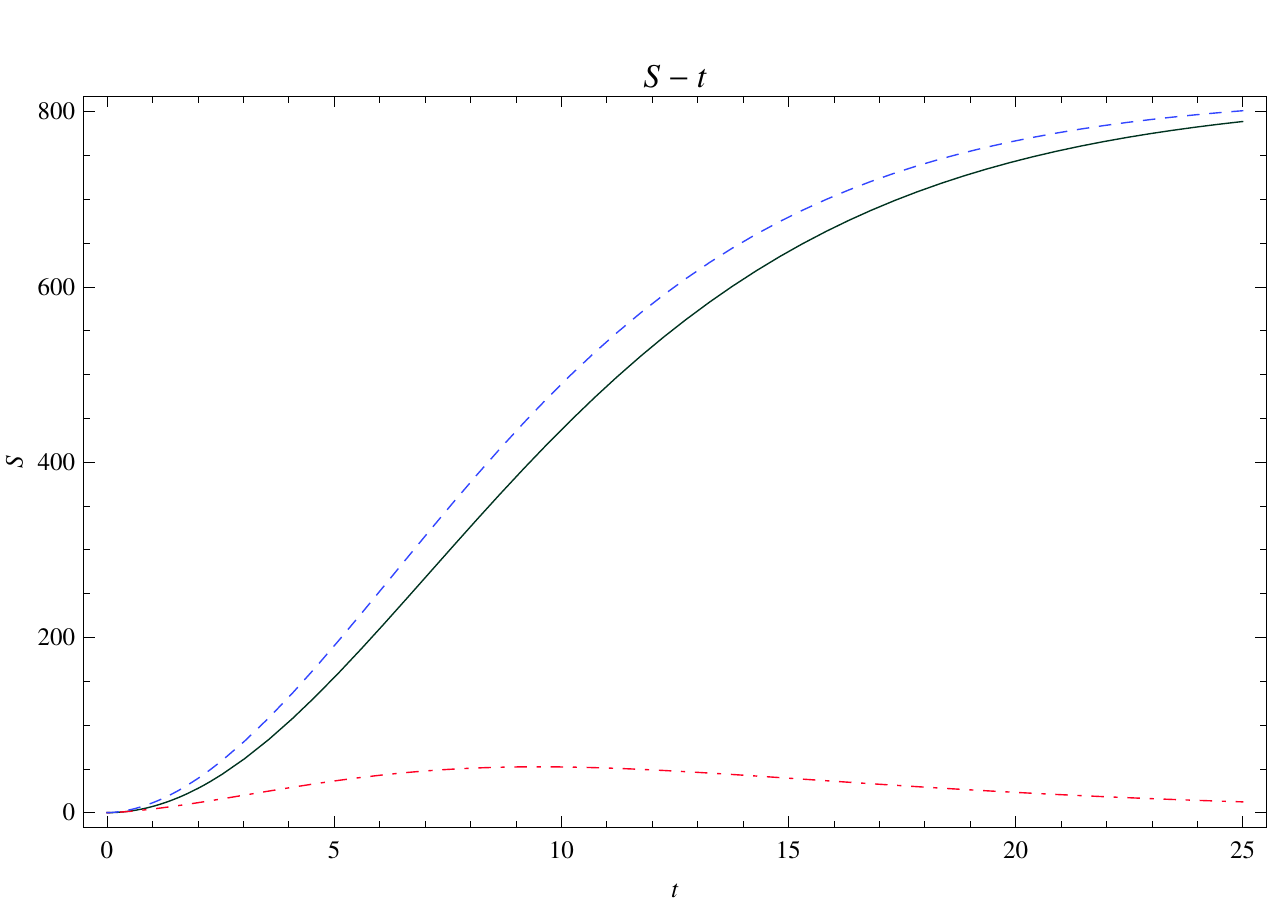}
 \caption{Variation $S_{CEH}. \, S_m$ and $S_{CEH}+S_m$ with time in Gyrs for Friedman universe with matter and 
cosmological constant. The continuous line representing entropy of horizon plus that of matter, 
dashed line representing the entropy of horizon and dash-dot line is for entropy of matter}
 \label{fig:totent2}
\end{figure}

Entropy of the event horizon is calculated as,
\begin{equation}
 S_{CEH} = {\pi \over H_0^2 \Omega_{\Lambda} \coth^2(\frac{3}{2}\sqrt{\Omega_{\Lambda}} H_0 t)}
\end{equation}
Entropy of matter can calculated using an analogous relation corresponds to equation (\ref{eqn:radentro}), and taking temperature 
of matter approximately as $\displaystyle T_m \sim \rho_m^{1/4}$, 
\begin{equation}
S_m  = {4\pi \over 3} {\rho^{3/4} \over H^3}  
\end{equation}
which after substitution of $H$ parameter become,
\begin{equation}
  S_m  = {4\pi \over 3} {\left(3 
H_0^2 \Omega_{\Lambda} \coth^2(\frac{3}{2}\sqrt{\Omega_{\Lambda}} H_0 t) - 3H_0^2 \Omega_{\Lambda} \right)^{3/4} 
\over H_0^3 \Omega_{\Lambda}^{3/2} \coth^3(\frac{3}{2}\sqrt{\Omega_{\Lambda}} H_0 t) }.  
\end{equation}
The behaviour of $S_{CEH} \, , \,  S_m$ and $S_{CEH}+S_m$ with time is shown in figure \ref{fig:totent2}.
The figure shows that the total entropy of the universe is increasing and the increase in the entropy of the horizon is more than that required for 
compensating the decrease in the matter entropy. The general behaviour is the same as that of Friedmann universe with radiation, that universe
with matter also will satisfy the generalized second law.

The condition for satisfying the generalized second law for this universe can be obtained by incorporating the time derivatives of 
the corresponding entropies into the second law, as
\begin{equation} \label{eqn:gslma}
 -{2\pi \dot{H} \over H} - \pi \rho_m^{3/4} \left( {3 \over H^2} + {4 \dot{H} \over H^4} \right) \geq 0
\end{equation}
Using the Hubble parameter equation (\ref{eqn:H2}), the above condition become,
\begin{equation}
\begin{split}
 {H_0 \sqrt{\Omega_{\Lambda}} Cosech^2(\frac{3}{2}\sqrt{\Omega_{\Lambda}}H_0 t ) \over Coth(\frac{3}{2}\sqrt{\Omega_{\Lambda}}H_0 t )}
- \rho_m^{3/4}  \\ \left(1 - 2 Sech^2(\frac{3}{2}\sqrt{\Omega_{\Lambda}}H_0 t ) \right) \geq 0
\end{split}
\end{equation}
Substituting $\rho_m$ in terms of the Hubble parameter using the Friedmann equation, we have plotted the time evolution of the left hand side
\begin{figure} 
 \includegraphics[scale=0.6]{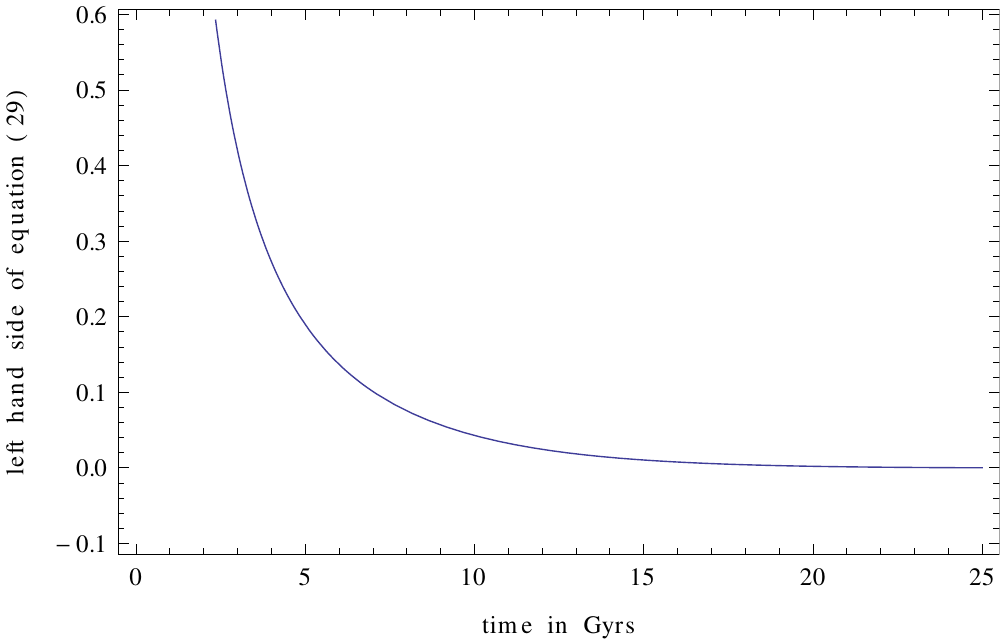}
\caption{Plot showing the validity of the generalized second law as per the condition in equation (29)}
\label{fig:secndlaw2}
\end{figure}
of the above equation and is shown in figure \ref{fig:secndlaw2}. As in the case of the Friedmann universe with radiation, here also the plot shows that 
the inequality condition corresponds to the generalized second law is satisfied.

As in the previous section, the generalized second law can leads to constraint on the temperature of matter. Equation (\ref{eqn:gslma})
can be recast, by taking $\dot{H} = -\rho_m/2$, as
\begin{equation}
 {\rho_m \over H} - \rho^{3/4} \left(3 - {2 \rho_m \over H^2} \right) \geq 0.
\end{equation}
From this it can be shown that, the present temperature of matter in the universe satisfies,
\begin{equation}
 T_{m0} \geq \left(3H_0 - 6 \Omega_{m0} H_0 \right)
\end{equation}
For the standard value $\Omega_{m0}=0.3$ the above condition also gives, $T_{m0} \geq 10^{-29} K.$ The temperature of the 
horizon is $T_h \sim H$ \cite{Gibbons1}, and with proper parameters, $T_h \sim 10^{-30} K.$ 
So the present temperature of the matter is greater than the temperature of the horizon, which supports 
the conclusion that the matter can cross the event horizon.

\section{Growth of event horizon and evolution of the entropy}

In this section we will restrict our analysis to Firedmann universe with radiation and cosmological constant only. However one can 
easily see that the conclusions made are in general true for Firedmann universe with matter also, but with different numerical
values. Our aim here is to show that the entropy of the contend of the universe does have a small increase before the development 
of event horizon.
In the last two sections we have discussed the behaviour of horizon entropy and entropy of the material within the horizon. We have 
concentrated on checking the validity of the generalized second law. We have
shown that the total entropy of the universe is always increasing, and the cosmological event horizon is satisfying the generalized
second law. However it is to be noted from the figure \ref{fig:entrorad} (from figure \ref{fig:totent2})
 that the entropy of the radiation (matter) is increasing first, 
attaining a maximum, then after it is decreasing. 
\begin{figure}
 \includegraphics[scale=0.6]{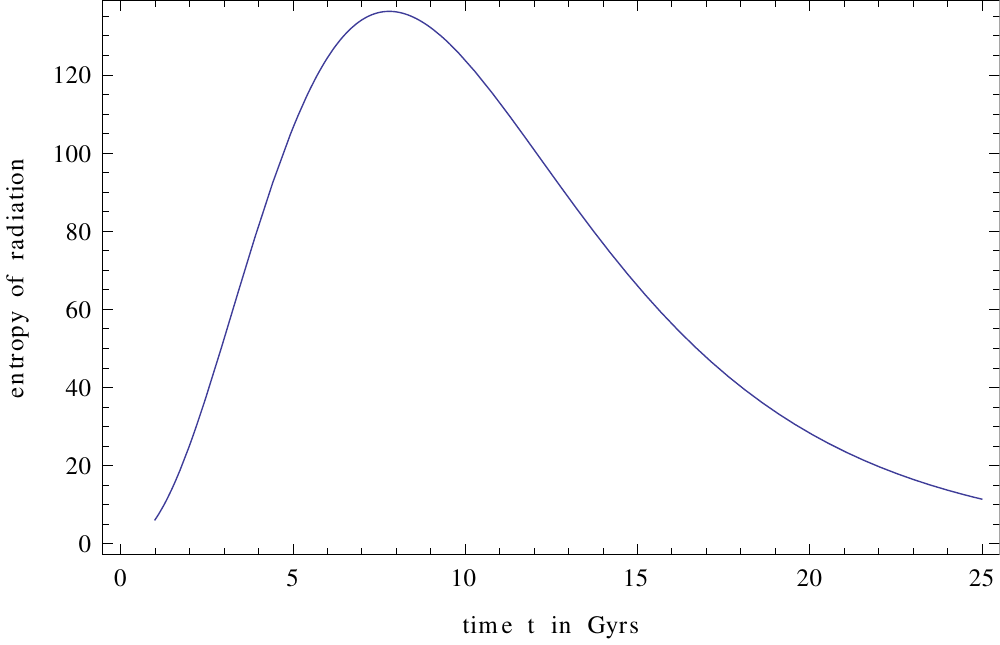}
\includegraphics[scale=0.6]{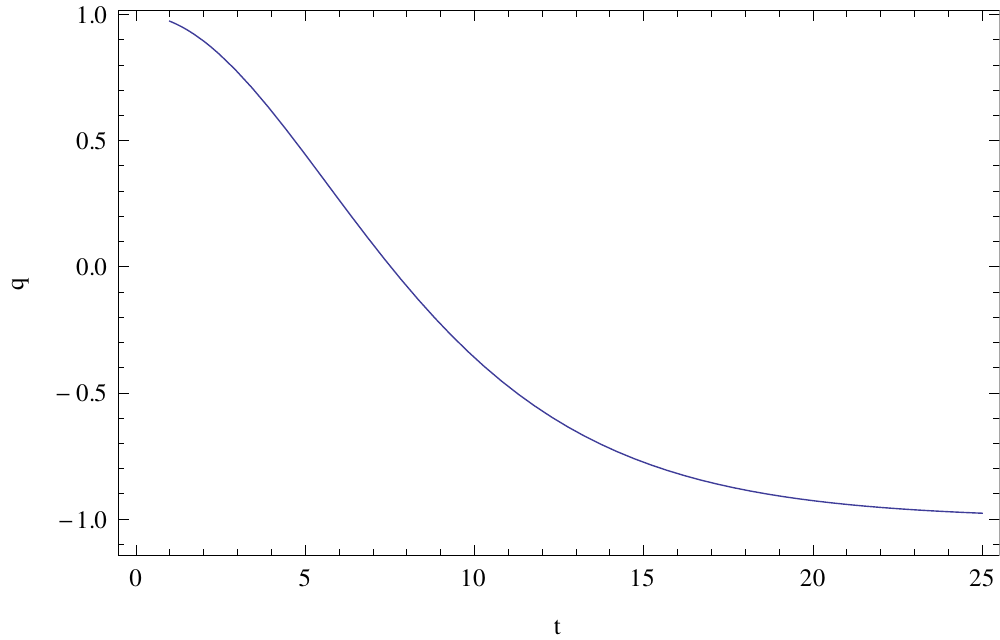}
\caption{The plot on the left represents the time evolution of the entropy of the radiation in the Friedmann universe with radiation 
and $\Lambda.$ and the plot on the right represents the time evolution of the $q-$factor of the same Friedmann universe with time in
giga years.}
\label{fig:radentro}
\end{figure}
For clarity regarding this we will show in figure \ref{fig:radentro}, the time evolution of the radiation 
entropy for a Friedmann universe having radiation and a positive cosmological constant. The figure shows that the radiation entropy 
first increases and then decreases to zero at very large times.

In the previous section we have concluded that the decrease in the entropy of radiation (or matter) is due to the escape of the radiation
(or matter) from within the horizon. The horizon will exist only when the universe is accelerating. From the bahaviour of the scale 
factor we have noted that, the universe will be in the radiation dominated (or matter dominated) phase as time $t \rightarrow 0.$ In the 
radiation dominated or matter dominated phase the expansion of the universe is decelerating, hence no horizon. The horizon 
will develop only when the universe enters the $\Lambda$ dominated phase. A clear demarcation between the deceleration and 
acceleration phases during the evolution of the universe can be obtained by calculating the deceleration parameter $q$, which is
defined as
\begin{equation}
 q = -1 - {\dot{H} \over H^2}
\end{equation}
we have calculated the $q-$factor using the Hubble parameter in equation (\ref{eqn:H1}) for the Friedmann universe with radiation and 
$\Lambda$ and the time evolution of which is shown in figure \ref{fig:radentro} along with the time evolution of the radiation entropy
for an easy comparison. The universe enters the accelerating phase, corresponding to the time at which
the $q-$factor starts to have negative values and as per the figure that is around a time $t \sim 7 Gyrs.$ 

As per the above analysis the universe enters the accelerating phase at around $t \sim 7 Gyrs$, and at around this time the event
horizon starts developing. At this transition time the horizon was tiny. Even at this time the difference in the entropy of the 
event horizon and radiation was very high. Entropy of the radiation given in the equation (\ref{eqn:sgamma2}), gives a value for 
$t= 7 Gyrs$, $S_{\gamma} \sim 10^{27}$. While the entropy of the event horizon, as in equation (\ref{eqn:CEH2}), leads to value of
$S_{CEH} \sim 10^{35}$, for the same time. These shows that, even at the formation of the event horizon, the entropy of it is eight 
orders of magnitude greater than the radiation entropy. So even at the tiny stage of the event horizon the entropy of radiation 
is not so significant.

A comparison of the plots in figure \ref{fig:radentro} shows that the entropy of radiation is increasing at first and 
is start decreasing at the same time when $q-$factor become negative, as the universe entering the accelerating phase. 
In the decelerating phase, corresponds to positive values of $q-$factor
the radiation entropy is increasing as there is no horizon for the radiation to escape. Since the radiation entropy is 
increasing during the initial stages, the generalized second law is still valid such that
\begin{equation}
 {dS_{\gamma} \over dt} \geq 0,
\end{equation}
 will become the GSL as there in no event horizon. When the 
universe enters the accelerated expanding phase, where it has event horizon, the radiation entropy is decreasing, because now the 
radiation is crossing the event horizon. But nevertheless, in the accelerating phase, the horizon entropy is increasing 
at a faster rate compensated to the decrease in the radiation entropy, which in turn leads to the increase in the total entropy 
of the universe, guaranteeing the validity of the generalized second law.

The time rate of radiation entropy is given in equation (\ref{eqn:radentrot}). 
Substituting for $\dot{H} = -2\rho_{\gamma}/3$ for Friedmann universe with radiation, the equation can be reduced to,
\begin{equation}
 {dS_{\gamma} \over dt} = {16\pi \sigma^{1/4} \over 3} \rho_{\gamma}^{1/4} \left({2\rho_{\gamma} \over 3 H^4} - {1 \over H^2}
\right)
\end{equation}
When the radiation entropy is maximum, the time rate is zero, then the above equation leads to the condition,
\begin{equation}
 H = \sqrt{2 \Lambda \over 3}
\end{equation}
where we have used the Friedmann equation to substitute for the $\rho_{\gamma}.$ From which the corresponding time can be 
obtained as
\begin{equation}
 t = {1 \over 2 \sqrt{\Omega_{\Lambda}} H_0} coth^{-1}(\sqrt{2})
\end{equation}
For the standard parameters, the value of the above time, corresponds to the decreasing of radiation
entropy, is around $t \sim 7 Gyrs,$ which is in confirmation with the figure \ref{fig:radentro}.
In the case Friedmann universe with matter and a positive cosmological constant also, it is evident form the figure \ref{fig:totent2}, 
that the entropy of matter too have an increase before the formation of the event horizon. So one can easily see that 
in the case matter also, the above conclusions are true in general.

\section{Conclusions}

Gibbons and Hawking have conjectured that cosmological event horizon of the de Sitter universe have entropy like black hole event
horizon, and the total 
entropy of such a universe will never decrease, that is it satisfies the GSL. Later Davies and others have extended this conjecture
to Friedmann universe with radiation and dust such that the Friedmann universe satisfies the GSL. However their work is mainly based on 
the numerical computation. In this paper we have presented an analytical analysis of the entropy of the event horizon and 
fluid within the horizon and also the constraints followed from the validity of the GSL. We 
have considered two types of Friedmann universes. Type one is the Friedmann universe with radiation and a positive cosmological constant.
The other type is the Friedmann universe with non-relativistic matter and a positive cosmological constant.

 We have obtained the expansion scale factor and the Hubble parameter for the Friedmann universe with radiation (and matter) and cosmological 
constant. The time evolution of the scale factor is plotted and have found that at sufficiently small times the Friedmann
universe is radiation(or matter) dominated and is in the decelerating phase. But at large times, the universe become dominated by 
the cosmological constant, hence in the accelerated expansion and will approach de Sitter phase at very large times. 
During the accelerated expansion phase, the universe has got an event horizon. We have numerically evaluated the time evolution of the 
comoving distance to the event horizon and verified that the comoving distance is decreasing with time in both types of the universes.
As a result the comoving volume of the event horizon decreases, subsequently the radiation (matter) can cross the event horizon.
This implies that the entropy of the radiation (or matter) is decreasing consequent to the escaping of radiation (or matter) through 
the horizon. 

Analogous to the area theorem in black hole, Davies proposed a corresponding theorem for the cosmological event horizon which 
implies a dominant energy condition as given in 
equation (\ref{eqn:condition1}). In the present case of the Friedmann universe, the dominant energy condition implies that, 
$H^2/ (H_0^2 \Omega_{\Lambda}) \geq 1.$ The plot in figure \ref{fig:cond1} conclusively proves this. So once the event horizon is
formed it's area will never decrease. So unlike in the case of the black holes, where the area of the event horizon decreases when it
is radiating, the area of the cosmological event horizon increases when radiation (matter) crosses the horizon.

We have obtained the analytical relations for the entropy of the event horizon and radiation (matter) for the Friedman universe. 
The entropy of the event horizon is given in equation (\ref{eqn:CEH2}), according to which the present value of the event horizon entropy
will be around, $S_{CEH} \sim  10^{35}.$ 
We have plotted the time evolution 
of these entropies and found that the net entropy of the radiation (or matter) is decreasing but the entropy of the event horizon is
increasing at faster rate as the universe expands. This implies that the total entropy of the Friedmann universe, that is the sum of the entropy of the radiation 
(or matter) and event horizon, is increasing. This indicate the validity of the GSL for both types of the Friedmann universes. The 
constraints imposed by the GSL is obtained. For the Friedmann universe with radiation and cosmological constant, the GSL 
constraint the present temperature of the radiation as, $T_{\gamma 0} \geq 10^{-29}$K in standard units. Compared to the latest value of the radiation 
temperature form COBE, 2.725$\pm$0.002 K \cite{Bennet1}, the above constraint implies the Friedmann universe in consideration 
is very well in the purview of the GSL. The temperature of the horizon, is $T_{CEH} = H$ in the standard units. For the 
the present case, this temperature is around $T_{CEH} \sim 10^{-30}$K. The comparison of the 
above temepratures shows that there is a radiation drain form within the horizon.
At this point one should note the result of Davies et. al.\cite{Davies2} that, the temperature of the radiation is higher than that of the horizon. 
So there is natural flow direction towards the horizon. Therefore in the present context, we can conclude that the temperature of the horizon
of the Friedmann universe with radiation and cosmological constant at present is less than $10^{-29}$ K. For the universe with 
non-relativistic matter and cosmological constant, the GSL constraint the matter temperature as $T_{m0} \geq 10^{-29}$K, which in turn 
implies that the horizon temperature of the Friedmann universe with matter and cosmological constant is less than $10^{-29}$K, so that
there is matter flow towards the event horizon.

The time evolution of the radiation (matter) entropy shows that, it increases first, attains a maximum and then decreases as shown 
in the figure \ref{fig:radentro}. It is seen that the increase in the radiation (matter) entropy is during the deceleration phase 
of the universe, when the radiation (matter) is dominating the cosmological constant. It is to be noted that there is no event horizon
when the universe is decelerating. So the corresponding increase in the entropy of the radiation is due to the non-existence of the
of the event horizon. If the event horizon is absent, there is no crossing of the radiation over the horizon, and it retained within
causal region of our universe, which facilitate the small increase in the radiation entropy. We made this point clear by comparing 
the time evolution of the radiation entropy and $q-$factor, such that the entropy of the radiation is start decreasing when the 
$q-$factor become negative, consequently the expansion is accelerating at which condition the universe posses an event horizon. We have
computed the the time corresponding to the maximum of the radiation entropy at which the $q-$factor is critically become negative, as
$t \sim 7Gyrs$, and is evident form figure \ref{fig:radentro}. 


\end{document}